\documentclass[aps,prl,preprint,notitlepage,groupedaddress,floatfix]{revtex4-2}

\usepackage{amsmath,amssymb,bm}
\usepackage{graphicx}
\usepackage{dcolumn}
\usepackage{amsthm}

\usepackage[utf8]{inputenc}
\usepackage{CJKutf8}
\newcommand{\zhname}[1]{\mbox{\begin{CJK}{UTF8}{gbsn}#1\end{CJK}}}

\numberwithin{equation}{section}

\newcommand{\dd}{\mathrm{d}}          % 正体微分元
\newcommand{\ii}{\mathrm{i}}          % 正体虚数单位
\newcommand{\ee}{\mathrm{e}}          % 正体自然指数

\begin{document}
	
	% ==================== 论文标题 ====================
	\title{Classical Mechanics Exactly Yields the Full Bound-State Spectrum of the Two-Dimensional Coulomb Problem}
	
	% ==================== 作者与单位 ====================
	\author{Gang Zheng \zhname{(郑罡)}$^{*}$}
	\author{Wenqi Xue \zhname{(薛文琪)}}
	\author{Mengli Wang \zhname{(王梦丽)}}
	\author{Peng Chen \zhname{(陈鹏)}}
	\author{Benniu Zhang \zhname{(张奔牛)}$^{*}$}
	\affiliation{%
		Chongqing Jiaotong University, No.66 Xuefu Avenue, Nan'an District, Chongqing, 400074, China \\
		\footnotesize $^{*}$\,Both authors are corresponding authors: zhenggang@cqjtu.edu.cn, benniuzhang@cqjtu.edu.cn
	}
	
	% ==================== 摘要 ====================
	\begin{abstract}
		High-lying Rydberg excitons in two-dimensional semiconductors universally exhibit a characteristic odd-integer energy scaling distinct from three-dimensional systems. While this hallmark of two-dimensional Coulomb interaction is well known from quantum mechanical solutions, its deeper classical geometric origin remains unclarified. Here we show that the complete bound-state spectral structure of the two-dimensional Coulomb problem---a central model for two-dimensional exciton physics---follows as an exact theorem from classical mechanics augmented by a single phase-scale parameter $\alpha$ with dimensions of action. We derive an amplitude-closure criterion as a necessary and sufficient condition for a classical propagator kernel to satisfy a linear evolution equation, and demonstrate that the singular Coulomb potential can be mapped shell-by-shell via Levi-Civita regularization into the class of quadratic Hamiltonians that obey this criterion exactly. The resulting spectrum bears odd-integer modal numbers, $1/N^2$ energy ratios and $N$-fold degeneracies, all independent of $\alpha$ and consistent with experimental observations of high-lying Rydberg excitons. This work provides a pure classical-geometry benchmark for two-dimensional exciton spectral analysis, allowing quantitative disentanglement of universal Coulomb effects from material-specific screening effects. No semiclassical, short-wavelength or $\hbar \to 0$ approximation is invoked at any stage. Our results invert the usual logical hierarchy for this integrable system: the wave equation emerges as a representation of the underlying classical geometry, rather than as an independent first principle.\\
		
		\textbf{keywords:}Two-dimensional Coulomb problem; Levi-Civita regularization; amplitude-closure criterion; classical-quantum correspondence; bound-state spectrum
	\end{abstract}

	\maketitle
	\newpage

	% ==================== 正文主体 ====================
	
	%----------------------------------section 1--------------------------------
	\section{Introduction}\label{sec:intro}
	
	The rise of van der Waals semiconductors such as monolayer transition-metal dichalcogenides has established two-dimensional excitons as a central research theme in condensed matter physics. Confined to an atomically thin plane and subject to weakened out-of-plane dielectric screening, excitons in these materials exhibit Coulomb binding energies of hundreds of meV and remain stable even at room temperature. Recent advances in optical spectroscopy have enabled clear observation of Rydberg exciton states up to principal quantum numbers $(n=6)$ and $(n=7)$ in materials such as WSe$_2$ and WS$_2$ \cite{19,20}. These high-lying states display a distinct two-dimensional spectral fingerprint: their energy levels asymptotically follow an odd-integer $1/(2n-1)^2$ scaling, in sharp contrast to the $1/n^2$ series of three-dimensional hydrogen. This characteristic Rydberg sequence is an intrinsic spectral signature of two-dimensional Coulomb interaction.
	
	In existing studies, this energy level structure is routinely obtained by directly solving the two-dimensional Schr\"odinger equation, and is treated as a standard result of quantum mechanics. However, a more fundamental question remains systematically unanswered: why does Coulomb interaction in two dimensions necessarily give rise to an odd-integer energy level sequence? Is this structure merely a mathematical solution of the quantum wave equation, or does it have a deeper root in classical geometry? This question also carries clear experimental relevance. Exciton interactions in real two-dimensional materials do not follow an ideal Coulomb potential, but instead obey a Rytova-Keldysh screened potential: the short-range regime is strongly modified by material polarization effects, while the long-range regime recovers Coulomb behaviour. Phenomenological hydrogen-like models are commonly used to fit experimental Rydberg spectra, but they cannot quantitatively distinguish which features are intrinsic universal properties of two-dimensional Coulomb interaction, and which arise from material-specific screening or lattice effects. A rigorous benchmark of the ideal two-dimensional Coulomb spectrum derived purely from classical mechanics would provide an entirely new frame of reference for quantitative analysis of experimental spectra.
	
	The correspondence between classical Hamilton-Jacobi theory and quantum mechanics forms one of the foundational themes of theoretical physics. Since the old-quantum-theory era, researchers have sought to interpret discrete energy levels from action integrals along classical orbits; the Sommerfeld-Einstein action-quantization rule first explained the discrete spectral structure of hydrogen based purely on classical orbital geometry \cite{1,2}. The subsequent Einstein-Brillouin-Keller (EBK) semiclassical quantization further generalized this scheme to non-separable integrable systems \cite{3}. In the standard narrative, quantum mechanics is taken as the fundamental description of nature, and classical mechanics is recovered in the limit of small de Broglie wavelength, or equivalently $\hbar \to 0$. Semiclassical methods, trajectory-based formalisms and path-integral techniques all operate within this framework: they employ classical phase-space information as a computational tool to approximate or reconstruct solutions to a given quantum wave equation \cite{4,5,6,7}.
	
	A long-standing open question lies outside this paradigm: can classical mechanics, by itself and without independent quantum axioms, exactly determine the discrete bound-state spectrum of a non-trivial system? Geometric quantization constructs quantum state spaces starting from classical symplectic manifolds \cite{8,9}, yet it fundamentally relies on quantization axioms via pre-quantum line bundles and does not escape the quantum-mechanical framework. For generic potentials the answer is almost certainly negative, but for highly integrable systems with special geometric structure, the possibility remains that spectral discreteness is a classical structural property, with the wave equation serving only as its linear representation.
	
	Previous work on classical-quantum mappings of the Coulomb problem, most notably the Duru-Kleinert transformation in path-integral quantization \cite{10,11}, employs classical regularization as a technical device within a postulated quantum framework. The Levi-Civita (LC) conformal map \cite{12}, originally developed to regularize two-body collision singularities in celestial mechanics, and its three-dimensional generalization, the Kustaanheimo-Stiefel transformation \cite{13,14}, are known to regularize the Coulomb singularity and map it to a harmonic oscillator, but this has been treated as a mathematical coincidence rather than a structural necessity. Recent attempts to reconstruct quantum propagators from ensembles of classical trajectories still presuppose the existence of quantum evolution equations and have triggered debate about their domain of validity \cite{15,16}. To date, no construction has derived the form of the linear evolution equation itself from classical consistency conditions, nor obtained the full discrete spectrum solely from classical phase-space data.
	
	In this Article, we present an exact classical construction of the two-dimensional Coulomb bound-state spectrum. Our approach rests on two pillars. First, we establish an amplitude-closure criterion: a classical kernel built from the two-point action and Van Vleck amplitude satisfies a linear Schr\"odinger-type evolution equation if and only if the spatial Laplacian of the amplitude vanishes identically. Second, we show that the LC conformal regularization maps every negative-energy Coulomb shell onto an isotropic harmonic oscillator in an image space---a quadratic Hamiltonian that falls exactly into the closure class. The even-subspace constraint imposed by the double-cover topology of the LC map then selects odd-integer modal numbers, yielding the characteristic Rydberg series, energy ratios and degeneracies in complete agreement with quantum results and experimental observations. All spectral ratios and degeneracy counts are independent of the phase-scale parameter $\alpha$, which sets only the absolute energy scale. The construction not only explains the structural necessity of the odd-integer Rydberg sequence from a classical perspective, but also provides a universal classical-geometry benchmark for quantitative separation of universal Coulomb effects and material-specific effects in experimental spectra.
	
	Our construction is exact and constructive at every step. It does not rely on semiclassical approximations, nor does it take the wave equation as given. For this integrable system, the discrete spectral structure is a direct consequence of classical phase-space geometry and the topology of the LC double cover.
	
	The remainder of this article is organized as follows. Section~\ref{sec:closure} develops the classical propagation kernel from first principles and proves the amplitude-closure criterion as a necessary and sufficient condition for exact equivalence between classical propagation and linear wave evolution. Section~\ref{sec:LC} introduces the Levi-Civita conformal regularization and demonstrates that the two-dimensional Coulomb problem maps shell-by-shell into the closure class of quadratic Hamiltonians. Section~\ref{sec:spectrum} imposes the topological boundary condition of the double-cover map and derives the full discrete bound-state spectrum together with its degeneracy structure. Section~\ref{sec:unique} establishes the uniqueness of the physical-space evolution operator and resolves the self-adjoint domain ambiguity at the origin from purely geometric considerations. Section~\ref{sec:discuss} discusses the distinction of this work from prior studies, its experimental implications for two-dimensional semiconductor excitons, and its foundational significance as well as open directions for generalization. Section~\ref{sec:conclusion} summarizes the principal conclusions.

	%----------------------------------section 2--------------------------------
	\section{Amplitude closure: exact equivalence between classical propagation and linear wave evolution}\label{sec:closure}
	
	%----------------------------------section 2.1------------------------------
	\subsection{Two-point action and the Van Vleck propagator kernel}\label{subsec:2.1}
	
	We begin by constructing a classical propagation kernel from first principles. For a Hamiltonian $H(q,p) = |p|^2/(2M) + V(q)$, the two-point classical action is defined as the integral of the Lagrangian along the unique classical orbit connecting source $q_0$ to endpoint $q$ in time $t$, valid in non-caustic regions of phase space:
	\begin{equation}\label{eq:2.1}
		S(q,t;q_0) = \int_0^t \mathcal{L}(\gamma, \dot{\gamma})\,\dd t',
	\end{equation}
	%------------------------equation 2.1
	where $\gamma$ denotes the classical trajectory satisfying $\gamma(0)=q_0$ and $\gamma(t)=q$. The endpoint momenta satisfy the standard Hamilton-Jacobi relations $p = \nabla_q S$ and $p_0 = -\nabla_{q_0} S$.
	
	The normalized Van Vleck amplitude describes the density of classical orbits transported by the Hamiltonian flow, given by the square root of the determinant of the mixed Hessian of $S$:
	\begin{equation}\label{eq:2.2}
		F = \sqrt{\det\left(-\frac{\partial^2 S}{\partial q_i \partial q_{0j}}\right)},
	\end{equation}
	%------------------------equation 2.2
	scaled by a dimensional constant consistent with Liouville measure transport \cite{4,17}.
	
	To construct a linear superposition representation of classical orbit ensembles, we introduce a phase-scale parameter $\alpha$ with dimensions of action, which converts the classical action into a dimensionless phase factor. We then construct a propagation kernel of the form
	\begin{equation}\label{eq:2.3}
		K_\alpha(q,t;q_0) = F(q,t;q_0)\,\ee^{\ii S(q,t;q_0)/\alpha},
	\end{equation}
	%------------------------equation 2.3
	where the numerical value of $\alpha$ is fixed by experimental calibration. As shown below, all spectral ratios and degeneracies are independent of its magnitude.

	%----------------------------------section 2.2------------------------------
	\subsection{Proof of the amplitude-closure equivalence criterion}\label{subsec:2.2}
	
	A central result of this work is that the classical kernel satisfies the linear evolution equation
	\begin{equation}\label{eq:2.4}
		\ii\alpha\partial_t K_\alpha = -\frac{\alpha^2}{2M}\nabla_q^2 K_\alpha + V(q) K_\alpha,
	\end{equation}
	%------------------------equation 2.4
	if and only if the amplitude closure condition
	\begin{equation}\label{eq:2.5}
		\nabla_q^2 F = 0
	\end{equation}
	%------------------------equation 2.5
	holds identically over all non-caustic regions of space and time.
	
	The proof proceeds by direct substitution. Computing the temporal and spatial derivatives of the kernel:
	
	\begin{align*}
		\partial_t K_\alpha &= \left(\partial_t F + \frac{\ii}{\alpha} F \partial_t S\right) \ee^{\ii S/\alpha},\\
		\nabla K_\alpha &= \left(\nabla F + \frac{\ii}{\alpha} F \nabla S\right) \ee^{\ii S/\alpha},\\
		\nabla^2 K_\alpha &= \left(\nabla^2 F + \frac{2\ii}{\alpha}\nabla F\cdot\nabla S + \frac{\ii}{\alpha}F\nabla^2 S - \frac{1}{\alpha^2}F|\nabla S|^2\right) \ee^{\ii S/\alpha}.
	\end{align*}
	
	Substituting into the linear evolution equation and cancelling the common exponential factor, the left-hand side becomes
	
	\begin{equation*}
		\ii\alpha \partial_t K_\alpha / \ee^{\ii S/\alpha} = \ii\alpha \partial_t F - F \partial_t S,
	\end{equation*}
	
	The right-hand side expands to
	
	\begin{equation*}
		-\frac{\alpha^2}{2M}\nabla^2 F - \frac{\ii\alpha}{M}\nabla F\cdot\nabla S - \frac{\ii\alpha}{2M}F\nabla^2 S 
		+ \frac{1}{2M}F|\nabla S|^2 + VF.
	\end{equation*}
	
	By the Hamilton-Jacobi equation $\partial_t S + \frac{1}{2M}|\nabla S|^2 + V = 0$, the terms of order $\alpha^0$ cancel identically on the real part of the equation. Collecting remaining terms and separating real and imaginary components, equality holds if and only if the real part and the imaginary part vanish separately. The imaginary part yields the closure condition $\nabla^2 F = 0$,
	while the real part reduces to the classical transport equation:
	
	\begin{equation}\label{eq:2.6}
		\partial_t F + \frac{1}{M}\nabla F\cdot\nabla S + \frac{1}{2M}F\nabla^2 S = 0.
	\end{equation}
	%------------------------equation 2.6
	The equivalence is therefore necessary and sufficient.
	
	Physically, this criterion states that the Van Vleck amplitude, which describes the density of classical trajectories, is a spatially harmonic function. This ensures that the interference pattern generated by the evolving orbit ensemble introduces no extra nonlinear terms, so that the ensemble evolution can be fully described by a linear wave equation.

	%----------------------------------section 2.3------------------------------
	\subsection{The closure class of quadratic Hamiltonians}\label{subsec:2.3}
	
	For quadratic Hamiltonians such as the isotropic harmonic oscillator, the two-point action is quadratic in the endpoints, so the Van Vleck amplitude depends only on time and not on position. The Laplacian of $F$ is therefore trivially zero, and the kernel satisfies the linear equation exactly---not approximately, not in a short-wavelength limit, but for all wavelengths and all times. For the two-dimensional isotropic harmonic oscillator with frequency $\omega$, the kernel takes the explicit form
	\begin{equation}\label{eq:2.7}
		K_\alpha(u,\tau;v_0) = \frac{M\omega}{-2\pi \ii\alpha \sin(\omega\tau)} 
		\times \exp\!\left[-\frac{\ii M\omega}{2\alpha\sin(\omega\tau)}\Bigl((r_u^2+|v_0|^2)\cos(\omega\tau)-2u\cdot v_0\Bigr)\right],
	\end{equation}
	%------------------------equation 2.7
	in which the amplitude prefactor depends solely on the time interval $\tau$, confirming that $\nabla_u^2 F = 0$ holds identically.
	
	Within the class of local potentials, quadratic Hamiltonians form the largest family that satisfies the closure condition exactly. Generic non-quadratic potentials cannot sustain $\nabla^2 F = 0$ globally, which is precisely why the Coulomb potential requires a regularization mapping to indirectly satisfy the closure criterion. Quadratic Hamiltonians thus form a closure class for which classical propagation and linear wave evolution are mathematically identical.
	
	This criterion establishes a general standard for deriving linear wave behaviour from classical mechanics, and points to a path for the exact solution of non-quadratic potentials such as the Coulomb problem: so long as the system can be mapped into the closure class, its wave evolution is entirely determined by classical geometry.

	%----------------------------------section 2.4------------------------------
	\subsection{Global properties of the kernel and caustic treatment}\label{subsec:2.4}
	
	The above results hold locally in non-caustic regions, where the two-point action is single-valued and the orbit connecting two endpoints is unique. At caustic surfaces, the map from initial momentum to final position degenerates, the Van Vleck amplitude diverges, and the single-valued description of $S$ breaks down.
	
	Globally, the Lagrangian submanifold of orbits retains a well-defined action; local patches are joined by Maslov phase factors, and the amplitude on each patch is given by the corresponding Van Vleck determinant. The kernel may be extended to a weak solution in the distributional sense across caustics, with temporal continuation implemented via Maslov patching. For test functions $\phi \in C_c^\infty(\mathbb{R}^2\times\mathbb{R})$, the kernel satisfies
	\begin{equation}\label{eq:2.8}
		\iint K_\alpha \left(-\ii\alpha\partial_t\phi - H^*\phi\right) \dd^2 q\,\dd t = 0
	\end{equation}
	in the distributional sense.
	%------------------------equation 2.8

	%----------------------------------section 2.5------------------------------
	\subsection{Field-theoretic symplectic formulation}\label{subsec:2.5}
	
	The linear evolution equation can also be cast in a real-symplectic field-theoretic framework, clarifying its relation to classical Hamiltonian flow. Decomposing the complex field $\phi = \phi_R + \ii\phi_I$, the field-theoretic symplectic form reads
	\begin{equation}\label{eq:2.9}
		\Omega_{\mathrm{field}}(X,Y) = \alpha\int (\phi_R \eta_I - \phi_I \eta_R)\,\dd^2 u,
	\end{equation}
	with Poisson bracket $\{\phi_R(u), \phi_I(u')\} = \alpha^{-1}\delta^2(u-u')$. 
	The corresponding Hamilton equations for the energy functional are mathematically equivalent to the complex linear evolution equation. In this sense, wavelike linear superposition corresponds to a complex representation of an underlying real-symplectic classical field flow.
	%------------------------equation 2.9

	%----------------------------------section 2.6------------------------------
	\subsection{Coherent and incoherent evolution of classical ensembles}\label{subsec:2.6}
	
	The kernel map can be extended to describe general classical ensembles characterized by an initial correlation kernel $\Gamma(q_0, q_0')$ encoding both amplitude weight and phase correlation between source points. The orbit-image intensity at position $q$ and time $t$ is defined as
	\begin{equation}\label{eq:2.10}
		\mathcal{I}_\Gamma(q,t) = \iint K_\alpha(q,t;q_0) K_\alpha^*(q,t;q_0') \, \Gamma(q_0,q_0') \,\dd^2 q_0 \,\dd^2 q_0'.
	\end{equation}
	%------------------------equation 2.10
	For a rank-one correlation kernel $\Gamma(q_0,q_0') = A(q_0) A^*(q_0')$, corresponding to a fully coherent initial ensemble, the intensity reduces to $|\phi(q,t)|^2$ with $\phi = T_\alpha A$, analogous to a pure quantum state. For a diagonal weight $\Gamma_{\mathrm{diag}}(q_0,q_0') = w(q_0) \delta^2(q_0 - q_0')$ describing an incoherent ensemble of independent classical sources, the intensity becomes
	\begin{equation}\label{eq:2.11}
		\mathcal{I}_{\mathrm{diag}}(q,t) = \int |K_\alpha(q,t;q_0)|^2 w(q_0) \,\dd^2 q_0,
	\end{equation}
	%------------------------equation 2.11
	which transports purely by Hamiltonian flow with no interference contributions. The difference between coherent and incoherent intensities defines the interference cross term, whose magnitude depends on the matching between the initial phase front and the classical action phase front.
	
	Despite the generality of the kernel framework, the Coulomb potential does not fall directly into the closure class due to its $1/r$ singularity and non-quadratic form. We show in the following section that a shell-wise regularization procedure maps every Coulomb bound-state shell precisely into this closure class.

	%----------------------------------section 3--------------------------------
	\section{Levi-Civita regularization maps the Coulomb problem into the closure class}\label{sec:LC}
	
	%----------------------------------section 3.1------------------------------
	\subsection{The conformal map and its double-cover topology}\label{subsec:3.1}
	
	The $1/r$ singularity of the Coulomb potential and its non-quadratic form prevent direct application of the amplitude-closure theorem. We address both issues via the LC conformal map, a holomorphic transformation originally developed to regularize two-body collision singularities in celestial mechanics \cite{12,18}.
	
	We define the physical complex coordinate $z = x + iy$ and the image-space coordinate $u$ via
	\begin{equation}\label{eq:3.1}
		z = \frac{u^2}{u_0},
	\end{equation}
	%------------------------equation 3.1
	where $u_0$ is a positive length scale. The conformal factor is
	\begin{equation}\label{eq:3.2}
		\lambda = \left|\frac{\dd z}{\dd u}\right|^2 = \frac{4r_u^2}{u_0^2} = \frac{4r_z}{u_0},
	\end{equation}
	%------------------------equation 3.2
	with $r_z = |z|$ and $r_u = |u|$.
	
\begin{figure}[htbp]
		\centering
		\includegraphics[scale=0.45]{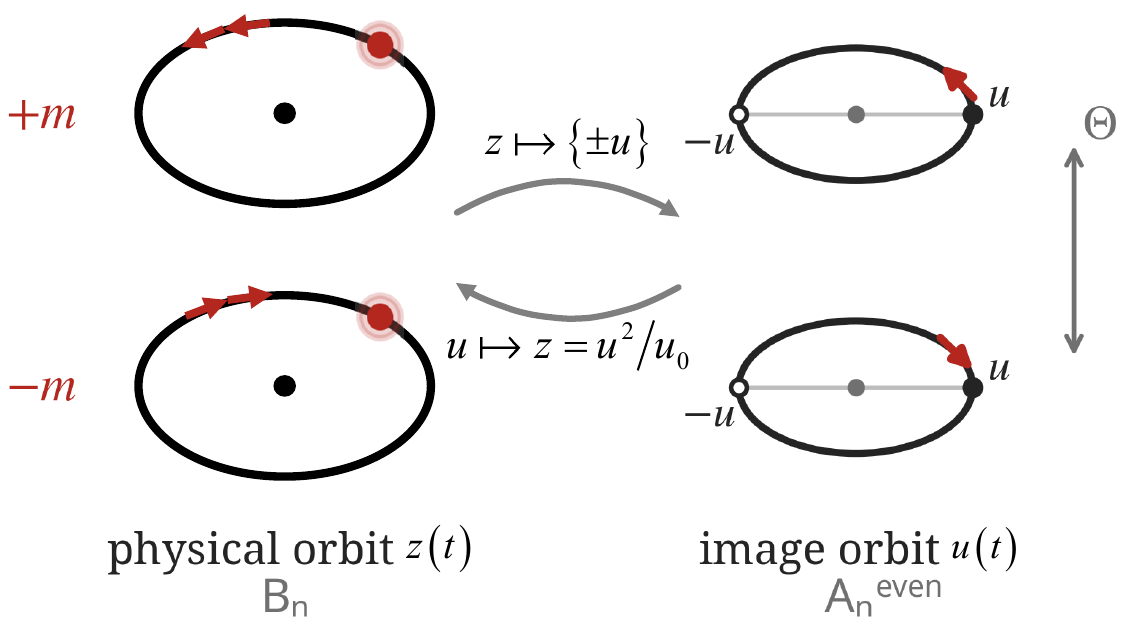}
		\caption{
			The LC chiral round trip. For a time-reversed pair $\pm m$, the physical orbit on the $n$-th bound-state shell lifts via $z\mapsto\pm u$ to an antipodal pair of the even image orbit, and maps back via $u\mapsto z=u^{2}/u_{0}$. Red chevrons mark the orbit chirality and evolution direction; the time-reversal operator $\Theta$ reverses the chirality and exchanges the two rows. The non-rotating $m=0$ mode is not shown.
			\label{fig1}
		}
	\end{figure}

	The map is a double cover of the physical $z$-plane: each point on the physical plane corresponds to two antipodal points $\pm u$ on the image plane. As illustrated in Fig.~\ref{fig1}, a single closed Kepler orbit in physical space lifts to a pair of antipodal oscillator orbits in image space, with opposite positions but identical chirality; the time-reversal operator $\Theta$ reverses the orbital chirality and exchanges the two antipodal branches. This topological structure directly gives rise to both the even-subspace constraint and the exact chirality correspondence of bound-state modes, and will impose a critical constraint on the physical spectrum derived in Section~\ref{sec:spectrum}.
	
	% ==================== 图1：LC手性往返示意图 ====================
	
	%----------------------------------section 3.2------------------------------
	\subsection{Shell-wise regularized Hamiltonian}\label{subsec:3.2}
	
	Since the Coulomb potential is non-quadratic and singular, the amplitude closure condition cannot be satisfied globally. For bound-state systems, however, each fixed energy corresponds to a closed invariant torus in classical phase space. We may therefore perform the regularization transformation shell by shell, mapping the motion on each energy shell to a quadratic Hamiltonian in the closure class.
	
	The conformal mapping rescales line elements, so that the kinetic term scales by $\lambda^{-1}$. To cancel the $1/r$ singularity of the Coulomb potential and convert it into a constant, we multiply the full Hamiltonian by the conformal factor $\lambda$, and subtract the fixed shell energy $E$ to obtain a time-independent Hamiltonian in image space. On any fixed negative-energy shell $H_z = E < 0$, we define the regularized Hamiltonian
	\begin{equation}\label{eq:3.3}
		H_{u,E} = \lambda(H_z - E).
	\end{equation}
	%------------------------equation 3.3
	The canonical one-form $p_z \cdot \dd z = p_u \cdot \dd u$ is preserved by the holomorphic map, ensuring that the image-space system is a genuine classical Hamiltonian system.
	
	Substituting the LC map term by term: the kinetic term transforms as $\lambda \cdot |p_z|^2/(2M) = |p_u|^2/(2M)$; the potential term gives $\lambda \cdot (-C/r_z) = -4C/u_0$, a constant offset independent of position; the energy term yields $\lambda \cdot (-E) = -4E r_u^2/u_0^2$, a quadratic restoring term.
	
	Combining these, we obtain
	\begin{equation}\label{eq:3.4}
		H_{u,E} = \frac{|p_u|^2}{2M} + \frac{1}{2}M\omega_E^2 r_u^2 - E_u,
	\end{equation}
	%------------------------equation 3.4
	where
	\begin{equation}\label{eq:3.5}
		\omega_E^2 = -\frac{8E}{M u_0^2}, \qquad E_u = \frac{4C}{u_0},
	\end{equation}
	%------------------------equation 3.5
	with $C>0$ the attractive Coulomb coupling constant.
	
	This is the Hamiltonian of an isotropic two-dimensional harmonic oscillator in the image space, with a shell-dependent frequency $\omega_E$ and a constant energy offset $E_u$. The length scale $u_0$ acts as a coordinate gauge in the LC representation: all physical predictions are independent of $u_0$, as changing $u_0$ rescales $u$, $\tau$, $\omega_E$, and $E_u$ jointly while leaving physical spectral values invariant.

	%----------------------------------section 3.3------------------------------
	\subsection{Structural uniqueness of the Coulomb--oscillator duality}\label{subsec:3.3}
	
	Crucially, this mapping is not an ad hoc trick. For the mapped system to strictly belong to the quadratic Hamiltonian closure class, the regularization transformation must simultaneously satisfy three structural requirements. First, potential flattening requires that $\lambda \cdot V(r_z)$ be constant, reducing the singular Coulomb potential to a uniform energy offset in image space. Second, energy-term quadraticity demands that $\lambda \cdot (-E)$ be proportional to $r_u^2$, yielding a quadratic restoring force characteristic of the harmonic oscillator. Third, kinetic standardization imposes that $\lambda = |\dd z/\dd u|^2$ with $z(u)$ holomorphic, preserving the conformal structure of the kinetic energy under coordinate transformation.
	
	The first condition implies $\lambda \propto r_z$, while the second implies $\lambda \propto r_u^2$. Combined with the third condition, this gives $|z'(u)|^2 \propto r_u^2$, whose unique holomorphic solution is $z(u) \propto u^2$. The three conditions are therefore mutually compatible if and only if the potential takes the Coulomb form and the map is quadratic holomorphic. The Coulomb potential is thus the unique central potential that can be shell-wise mapped into a quadratic Hamiltonian via holomorphic conformal regularization. The Coulomb--oscillator duality is a structural necessity, not a mathematical coincidence\cite{21}.

	%----------------------------------section 3.4------------------------------
	\subsection{Integrability of the regularized oscillator}\label{subsec:3.4}
	
	As a quadratic rotationally invariant Hamiltonian, the image-space oscillator is completely integrable in the Liouville-Arnold sense. The Hamiltonian $H_{u,\mathrm{osc}} = |p_u|^2/(2M) + \frac{1}{2}M\omega_E^2 r_u^2$ is conserved, as is the angular momentum $L_u = u_x p_{u,y} - u_y p_{u,x}$ by rotational symmetry. Their Poisson bracket vanishes identically, and their gradients are linearly independent almost everywhere. With two degrees of freedom and two independent commuting conserved quantities, the system is fully integrable, ensuring that motion is confined to invariant tori in phase space.

	%----------------------------------section 3.5------------------------------
	\subsection{Layered complex structures}\label{subsec:3.5}
	
	Four mathematically distinct complex structures appear in this construction, each operating on a distinct space and logically independent of one another. The first is the configuration complex structure of multiplication by $i$ in the image plane $u = u_x + iu_y$, which encodes spatial rotations in two dimensions. The second is the single-particle phase-space symplectic structure that mixes position and momentum variables, underlying the Hamiltonian flow. The third is the field-theoretic phase-space complex structure mixing the real and imaginary components of the field, which gives rise to the linear superposition principle in the wave representation. Finally, the scalar complex unit $i$ in the phase factor $\ee^{\ii S/\alpha}$ sets the oscillatory scale of the kernel. Distinguishing these layers avoids conceptual confusion between geometric rotations, symplectic structure, and wave phase.
	
	As a quadratic Hamiltonian, the image-space oscillator belongs exactly to the amplitude-closure class: its Van Vleck kernel satisfies the linear evolution equation without approximation. With each energy shell mapped to an image-space harmonic oscillator, we now impose the physical boundary conditions and derive the discrete spectrum.

	%----------------------------------section 4--------------------------------
	\section{Odd-integer Rydberg series from the even-subspace constraint}\label{sec:spectrum}
	
	Since the regularized classical system satisfies the linear evolution equation~(\ref{eq:2.4}), its stationary solutions correspond to time-separable bound-state modes, governed by an eigenvalue equation with the same generator as the evolution equation. We may therefore derive the complete bound-state spectrum of the physical system directly from the eigenproblem of the image-space oscillator, combined with the topological constraints of the physical space.
	
	%----------------------------------section 4.1------------------------------
	\subsection{Eigenstates of the image-space harmonic oscillator}\label{subsec:4.1}
	
	The image-space isotropic harmonic oscillator has well-known eigenstates obtained via separation of variables in polar coordinates. Writing $\Psi_u(u) = f(r_u) \ee^{\ii m_u\theta_u}$, the radial eigenvalue equation reads
	\begin{equation}\label{eq:4.1}
		-\frac{\alpha^2}{2M}\left[f'' + \frac{1}{r_u}f' - \frac{m_u^2}{r_u^2}f\right] + \frac{1}{2}M\omega_E^2 r_u^2 f = E_u f,
	\end{equation}
	%------------------------equation 4.1
	where $n_r = 0,1,2,\dots$ is the radial node number and $m_u \in \mathbb{Z}$ is the image-space angular momentum quantum number.
	
	Imposing the decay boundary condition $f(r_u) \sim \exp\bigl(-M\omega_E r_u^2/(2\alpha)\bigr)$ as $r_u \to \infty$ and the regularity condition $f(r_u) \sim r_u^{|m_u|}$ as $r_u \to 0$, the energy eigenvalues are
	\begin{equation}\label{eq:4.2}
		E_u = \alpha\omega_E\bigl(2n_r + |m_u| + 1\bigr).
	\end{equation}
	%------------------------equation 4.2
	This result is exact for all quantum numbers, with no semiclassical approximation invoked.

	%----------------------------------section 4.2------------------------------
	\subsection{Even-subspace constraint from the double-cover topology}\label{subsec:4.2}
	
	The double-cover topology of the LC map implies that a closed classical orbit in physical space corresponds to an orbit that completes two circuits in image space. For the propagation kernel and all physical observables to be single-valued on the physical plane, two image-space points separated by a $2\pi$ rotation must correspond to the same physical state. Only image-space modes with even angular momentum therefore correspond to physically self-consistent classical motion. This even-subspace constraint restricts the image angular momentum to even values: $m_u = 2m$, where $m \in \mathbb{Z}$ is the physical angular momentum index.

	%----------------------------------section 4.3------------------------------
	\subsection{Physical bound-state energy spectrum}\label{subsec:4.3}
	
	Substituting $m_u = 2m$ into the oscillator energy condition and defining the odd-integer modal number
	\begin{equation}\label{eq:4.3}
		N = 2n_r + 2|m| + 1 = 2n - 1, \quad n = 1,2,3,\dots,
	\end{equation}
	%------------------------equation 4.3
	we derive the physical bound-state energy spectrum. From the eigenvalue condition $E_u = \alpha\omega_E N$ we solve for $\omega_E = E_u/(\alpha N)$, and substitute into the definition of $\omega_E^2$ to obtain the physical energy:
	\begin{equation*}
		E = -\frac{Mu_0^2 \omega_E^2}{8} = -\frac{Mu_0^2}{8} \left( \frac{E_u}{\alpha N} \right)^2,
	\end{equation*}
	Inserting $E_u = 4C/u_0$, the $u_0$ gauge cancels out identically, yielding
	\begin{equation}\label{eq:4.4}
		E_n = -\frac{2MC^2}{\alpha^2 N^2} = -\frac{2MC^2}{\alpha^2 (2n-1)^2}.
	\end{equation}
	%------------------------equation 4.4
	The spectral ratios follow immediately:
	\begin{equation}\label{eq:4.5}
		\frac{E_n}{E_1} = \frac{1}{(2n-1)^2},
	\end{equation}
	%------------------------equation 4.5
	yielding the characteristic sequence $1 : 1/9 : 1/25 : \cdots$ that distinguishes two-dimensional Coulomb dynamics from its three-dimensional counterpart.

	%----------------------------------section 4.4------------------------------
	\subsection{Degeneracy structure and chirality correspondence}\label{subsec:4.4}
	
	The degeneracy of the $n$-th shell is $g_n = 2n-1$, comprising one non-rotating $(m = 0)$ mode and $n-1$ time-reversed pairs of opposite chirality ($\pm m$). This degeneracy structure is entirely determined by the topology of the LC double cover and the integrability of the classical oscillator, with no quantum axioms required.
	
	There is an exact correspondence between classical orbit chirality and modal chirality. For the image-space oscillator, the classical angular momentum $L_u$ satisfies $L_u = \alpha m_u$ exactly, so $\operatorname{sgn}(L_u) = \operatorname{sgn}(m_u)$. Under the LC map, $L_u = 2L_z$ and $m_u = 2m$, so $\operatorname{sgn}(L_u) = \operatorname{sgn}(m)$ transfers to physical-space modes. Modal chirality, defined by the circulation direction of the probability current $j_{\theta_z} = (\alpha m / (M r_z)) |\Psi_m|^2$, thus matches the orientation of the corresponding classical orbital angular momentum exactly for all $\alpha>0$.

	%----------------------------------section 4.5------------------------------
	\subsection{Winding number dynamics and phase periodicity}\label{subsec:4.5}
	
	The classical geometric origin of the discrete spectrum can also be understood intuitively from the winding number of the action along closed orbits. The energy quantization condition is equivalent to the requirement that the action increment on a closed orbit is an integer multiple of the phase scale, which is determined entirely by the topology of classical phase space.
	
	For an image-space harmonic oscillator orbit, the increment of the reduced action per image period is
	\begin{equation}\label{eq:4.6}
		\Delta S_{\mathrm{red}} = \oint p_u\cdot\dd u = \frac{2\pi E_u}{\omega_E}.
	\end{equation}
	%------------------------equation 4.6
	Using $E_u = \alpha\omega_E N$, this gives $\Delta S_{\mathrm{red}} = 2\pi\alpha N$,
	corresponding to a winding-number increment $\Delta\nu = N$,
	per image period.
	
	One image period corresponds to two physical Kepler periods, so the reduced-action increment per physical period is $\Delta S_{\mathrm{red}} = \pi\alpha N$,
	yielding a spatial phase factor $\ee^{\ii\pi N} = -1$,
	since $N$ is odd. The Maslov phase from two radial turning points contributes $\ee^{-\ii\pi} = -1$,
	and their product is $+1$, ensuring single-valuedness of the stationary-wave phase on the closed orbit. This phase matching is a direct consequence of classical phase-space topology, with no independent quantum postulate needed.

	%----------------------------------section 4.6------------------------------
	\subsection{The phase-scale parameter $\alpha$}\label{subsec:4.6}
	
	All spectral ratios, degeneracy counts, and modal structures derived above are independent of the phase-scale parameter $\alpha$, which sets only the absolute energy scale. The parameter $\alpha$ carries dimensions of action, identical to the dimension of Planck's constant $\hbar$, but its functional role differs fundamentally: rather than entering as an independent fundamental postulate, $\alpha$ is a free phenomenological parameter calibrated by experimental data.
	
	For a given system, if the ground-state binding energy $|E_1|$ is measured, $\alpha$ is determined via
	\begin{equation}\label{eq:4.7}
		\alpha = \frac{\sqrt{2M}\,C}{|E_1|^{1/2}}.
	\end{equation}
	%------------------------equation 4.7
	Once calibrated, the full Rydberg series and degeneracy structure are predicted with no further adjustable parameters. This work makes no ontological claim about the relation between $\alpha$ and $\hbar$; the structural properties of the spectrum are determined by classical geometry alone, with $\alpha$ serving only as an overall scaling factor.
	
	The above derivation confirms that the discreteness of the spectrum, the energy level ratios, and the degeneracy structure are fully determined by the topological properties of the LC map and the integrability of the classical oscillator, with no independent quantization postulates introduced. Beyond reproducing the full bound-state spectral structure, our construction also uniquely determines the form of the local evolution operator on the physical plane, and resolves the long-standing ambiguity of self-adjoint extensions at the origin from purely geometric considerations. These results are presented in the next section.

	%----------------------------------section 5--------------------------------
	\section{Unique Hamiltonian and self-adjoint domain from classical geometry}\label{sec:unique}
	
	Beyond reproducing the spectrum, our construction uniquely determines the form of the local evolution operator on the physical plane, and resolves the subtle issue of self-adjoint extensions at the origin entirely from geometric considerations.
	
	%----------------------------------section 5.1------------------------------
	\subsection{Pullback identity of the differential operator}\label{subsec:5.1}
	
	The amplitude-closure criterion has already established that the generator of the classical propagation kernel is a second-order linear differential operator of the form~(\ref{eq:2.4}). The classical regularization mapping from physical space to image space therefore also uniquely determines the form of the physical-space evolution operator.
	
	We begin from the conformal transformation rule for the two-dimensional Laplacian under the holomorphic LC map. With the conformal factor $\lambda = |\dd z/\dd u|^2$, the Laplacian on the physical plane satisfies
	\begin{equation}\label{eq:5.1}
		\nabla_z^2 = \frac{1}{\lambda}\nabla_u^2,
	\end{equation}
	%------------------------equation 5.1
	pointwise on the punctured plane $u \neq 0$.
	
	Defining the image-space oscillator operator
	\begin{equation}\label{eq:5.2}
		H_{u,\mathrm{osc}} = -\frac{\alpha^2}{2M}\nabla_u^2 + \frac{1}{2}M\omega_E^2 r_u^2
	\end{equation}
	%------------------------equation 5.2
	and the physical-space operator
	\begin{equation}\label{eq:5.3}
		H_z = -\frac{\alpha^2}{2M}\nabla_z^2 - \frac{C}{r_z},
	\end{equation}
	%------------------------equation 5.3
	one directly verifies the shell-wise identity
	\begin{widetext}
		\begin{equation}\label{eq:5.4}
			\lambda(H_z - E)\Psi = (H_{u,\mathrm{osc}} - E_u)\Psi_u,
		\end{equation}
		%------------------------equation 5.4
	\end{widetext}
	where $\Psi_u(u) = \Psi(z(u))$ is the pure pullback of the physical field to image space. This identity holds for every fixed negative-energy shell, and underpins the exact equivalence between the two representations.
	
	Three distinct representational layers arise in conjunction with this mapping. At the equation layer, the pure pullback preserves the shell-wise differential identity exactly, with no additional factors. At the normalization layer, shell-dependent constant factors relate normalized image modes to normalized physical modes; for an image-space eigenmode of unit norm, the physical-space norm involves the expectation value $\langle r_u^2\rangle_n$, fixed by the virial theorem for the harmonic oscillator. At the measure layer, multiplication by $\lambda^{-1/2}$ preserves the Hilbert inner product under coordinate transformation, but unitary conjugation by $\sqrt{\lambda}$ introduces gradient and geometric corrections to the local differential operator. The physical-space local generator is therefore determined by the equation-layer pure-pullback identity, not by measure transport.

	%----------------------------------section 5.2------------------------------
	\subsection{Uniqueness theorem for the local evolution operator}\label{subsec:5.2}
	
	The above construction not only recovers the standard Coulomb Hamiltonian, but fixes its form uniquely from the requirement of shell-wise equivalence to the regularized image-space oscillator. We state this result as a theorem:
	
	\textit{Theorem.} Let $\mathcal{D} = -a\nabla_z^2 + V_T(z)$ be any second-order local differential operator on $z \neq 0$ with constant $a>0$ and real potential $V_T$. If, for every $E<0$, transplantation to image space yields the regularized-oscillator eigenvalue equation with the corresponding shell-dependent frequency $\omega_E$ and fixed offset $E_u$, then
	\begin{equation}\label{eq:5.5}
		a = \frac{\alpha^2}{2M}, \qquad V_T(z) = -\frac{C}{r_z}.
	\end{equation}
	%------------------------equation 5.5
	\textit{Proof.} Matching the second-derivative coefficient using the conformal Laplacian relation $\nabla_z^2 = \lambda^{-1}\nabla_u^2$ fixes the kinetic prefactor $a = \alpha^2/(2M)$ uniquely. Matching the $E$-independent constant terms in the shell identity fixes the potential to the Coulomb form $V_T = -C/r_z$. The two-dimensional conformal Laplacian admits no first-order correction term, so no additional gradient or vector-potential term appears in the physical-space operator. Both the kinetic coefficient and the Coulomb potential are thus fixed uniquely by the requirement of shell-wise compatibility with the image-space oscillator. No free parameters remain in the local differential expression once the classical map is specified. \hfill$\blacksquare$

	%----------------------------------section 5.3------------------------------
	\subsection{Self-adjoint domain at the origin}\label{subsec:5.3}
	
	The operator identity holds pointwise on the punctured plane $z \neq 0$; the behaviour at the origin requires separate treatment via self-adjoint extension theory. After standard radial reduction $y = r_z^{1/2}\psi$, the effective inverse-square centrifugal coefficient takes the form $m^2 - 1/4$.
	
	For $|m| \geq 1$, the coefficient is greater than or equal to $3/4$, placing the origin in the limit-point regime. The self-adjoint extension is unique, with no free boundary condition at the force centre. For $m = 0$, the effective coefficient is exactly $-1/4$, the critical coupling for fall-to-the-centre. This channel falls into the limit-circle regime: both the regular solution $\psi \sim r_z^{1/2}$ and the singular solution $\psi \sim r_z^{1/2}\ln r_z$ are square-integrable at the origin, admitting a one-parameter family of mathematically valid self-adjoint extensions.
	
	In standard quantum mechanics, the regular (Friedrichs) extension is selected on physical grounds, but without a first-principles justification. Our construction resolves this ambiguity geometrically. From the perspective of classical mechanics, the harmonic oscillator potential in image space is everywhere regular. Classical orbits never develop infinite velocity or divergent ensemble density at the origin, so the corresponding solution must be analytic and finite there. The regular solution with analytic eigenfunctions at the origin is therefore the natural and unique choice, with no additional boundary condition freedom. Pulling this domain back to physical space via the LC map automatically selects the regular branch for the $m = 0$ channel, yielding wavefunctions of the form
	\begin{equation}\label{eq:5.6}
		\Psi_m(r_z) = r_z^{|m|} f_m(r_z), \qquad f_m(0)\ \text{finite},
	\end{equation}
	%------------------------equation 5.6
	and excluding the logarithmic singular solution. The boundary condition at the force centre is therefore not an independent postulate, but a consequence of the regular classical structure of the image space.

	%----------------------------------section 5.4------------------------------
	\subsection{Completeness of the bound-state basis}\label{subsec:5.4}
	
	For the regular self-adjoint extension selected by the LC construction, the physical-space Hamiltonian satisfies three key spectral properties. First, its essential spectrum is $\sigma_{\mathrm{ess}}(H_z) = [0,\infty)$, corresponding to positive-energy scattering states. Second, the negative-energy spectrum consists entirely of the discrete eigenvalues derived in Section~\ref{sec:spectrum}, with no additional bound states. Third, the corresponding eigenfunctions form a complete orthonormal basis of the negative-spectral subspace.
	
	The first property follows from relative compactness of the Coulomb potential with respect to the kinetic energy operator. The second follows from explicit solution of the radial equation in each angular channel, which exhausts all square-integrable regular solutions. The third is then a direct consequence of the spectral theorem for self-adjoint operators. The complete bound-state spectral problem is thus fully determined by classical geometric data.
	
	Having established the uniqueness of the evolution operator and its self-adjoint domain on purely geometric grounds, we turn in the following section to a broader discussion of the results, their relation to prior work, and their experimental and foundational implications.

	%----------------------------------section 6--------------------------------
	\section{Discussion}\label{sec:discuss}
	
	%----------------------------------section 6.1------------------------------
	\subsection{Distinction from prior work}\label{subsec:6.1}
	
	The core difference between this work and all existing studies lies not in computational technique, but in logical hierarchy. All prior classical-quantum correspondence studies take the quantum wave equation as a first principle, and use classical mechanics as an approximation or computational tool. In this work, by contrast, the linear wave equation is a linear representation of the underlying classical geometric structure, and is thus a derived result.
	
	Semiclassical quantization, WKB methods and the Einstein-Brillouin-Keller action quantization are asymptotic procedures that yield approximate spectra for large quantum numbers \cite{1,2,3}. Geometric quantization builds quantum Hilbert spaces from symplectic geometry yet imposes quantization rules externally via pre-quantum line bundles, and does not escape the quantum-mechanical framework \cite{8,9}. The Duru-Kleinert transformation regularizes the Coulomb path integral, but it is formulated entirely within a quantum-mechanical framework where the propagator and the evolution equation are taken as given \cite{10,11}.
	
	By contrast, we derive the linear evolution equation itself from a classical consistency condition---the amplitude-closure criterion---and obtain the spectrum by imposing classical topological constraints. The amplitude-closure criterion is an exact structural condition, not a truncated asymptotic expansion; the LC-regularized image-space kernel satisfies the linear equation exactly, not approximately. The results of this work therefore do not belong to a semiclassical, short-wavelength, or $\hbar \to 0$ asymptotic regime. The construction is exact and constructive at every step, and holds for all quantum numbers including the ground state, in contrast to semiclassical methods which are valid only in the large-quantum-number limit.
	
	The key conceptual shift is that the wave equation is not taken as a starting point. It emerges as the natural linear representation of a classical system whose transported density amplitude satisfies the harmonicity condition. For the class of closed quadratic Hamiltonians, and for any system that can be shell-wise regularized into that class, wavelike behaviour is an exact property of classical ensemble dynamics.

	%----------------------------------section 6.2------------------------------
	\subsection{Experimental implications for two-dimensional excitons}\label{subsec:6.2}
	
	Our results provide a pure classical-geometry benchmark for interpreting exciton spectra in two-dimensional semiconductors. Real materials exhibit Rytova-Keldysh-type screened potentials that deviate from pure $1/r$ behaviour at short distances \cite{22,23}, causing the low-lying $1s, 2s$ states to depart from the ideal Rydberg series. High-lying Rydberg states, by contrast, sample predominantly the long-range part of the potential and should approach the odd-integer $1/N^2$ series predicted here.
	
	The classical benchmark can thus be used to quantitatively disentangle universal Coulombic effects from material-specific screening and lattice effects in experimental spectra. Deviations from the $1/(2n-1)^2$ scaling at different principal quantum numbers provide a direct measure of the range and strength of non-ideal screening.
	
	Practically, this framework enables two key applications in experimental spectral analysis. First, it allows quantitative extraction of screening strength: by comparing experimentally measured high-lying energy levels to the classical benchmark, one can directly infer the effective dielectric screening strength of the material, and resolve the modulation of exciton binding energy by different substrates or encapsulation structures. Second, it enables a clean separation between universal and material-specific effects: odd-integer energy level ratios and degeneracy structures are universal properties of two-dimensional Coulomb interaction, while low-level energy shifts, valley splitting and spin splitting are material-specific effects. With the classical spectrum as a reference, the two contributions can be clearly distinguished, avoiding the misidentification of universal geometric effects as novel material properties.
	
	Compared with traditional quantum hydrogen-like models, this classical benchmark offers a clearer physical attribution: all features that follow the odd-integer scaling are essentially classical geometric products of two-dimensional Coulomb interaction, independent of the quantization mechanism.
	
	The absolute energy scale is set by the phase parameter $\alpha$, which may be calibrated directly from experimental data. If the ground-state binding energy $|E_1|$ is measured, $\alpha$ is determined via Eq.~(\ref{eq:4.7}). If $M$ and $C$ are known from independent measurements, this relation fixes $\alpha$ uniquely; if they are effective material parameters, the combination $2MC^2/\alpha^2$ is fixed first by experiment, with separate determinations of effective mass and coupling then yielding $\alpha$ individually. In either case, all spectral ratios and degeneracies are invariant under changes in $\alpha$, and are predicted with no further adjustable parameters once the system is calibrated.

	%----------------------------------section 6.3------------------------------
	\subsection{Logical hierarchy and foundational significance}\label{subsec:6.3}
	
	The full construction is organized into three logically distinct layers, each with well-defined mathematical status. At the inner classical layer reside the symplectic structure, two-point action, Lagrangian submanifolds, Liouville measure, and the LC conformal factor---purely classical mechanical objects with no reference to linear superposition or wave behaviour. At the representation layer, the phase scale $\alpha$ is introduced, the kernel function is constructed, and homotopic orbit families are classified, building a linear representation of the underlying classical data. At the spectral conclusion layer, the amplitude-closure criterion, the quadratic closure theorem, the shell-wise LC equivalence, the local-operator uniqueness, and the discrete spectral structure are derived entirely from the classical data of the inner layer.
	
	It is worth clarifying that complex phases and linear superposition are formal tools adopted at the representation layer, not independent physical postulates. In this work, all core physical properties that determine the spectrum---discreteness, energy level ratios, degeneracy counts, and chirality decomposition---are uniquely fixed by the topology and geometry of classical phase space, independent of the choice of representation. The wave equation serves only as a linear encoding of these classical properties.
	
	At a foundational level, our results invite a re-evaluation of the logical hierarchy between classical and quantum mechanics for integrable systems. That the full discrete spectrum of a non-trivial bound-state problem can be obtained from classical geometry suggests that quantum discreteness, at least in highly symmetric systems, may be less mysterious than often supposed---it may simply reflect the topology of classical phase space encoded in a linear representation. For this integrable system, the wave equation emerges as a representation of the underlying classical geometry, rather than as an independent first principle.

	%----------------------------------section 6.4------------------------------
	\subsection{Limitations and outlook}\label{subsec:6.4}
	
	Several important open directions remain. First, the present work is restricted to negative-energy bound states with decaying boundary conditions at infinity. Extending the construction to the positive-energy continuous spectrum and scattering states is a natural next step, and will require careful treatment of boundary conditions at infinity as well as self-adjoint extension properties of the positive-energy sector.
	
	Second, generalization to three-dimensional hydrogen via the Kustaanheimo-Stiefel transformation \cite{13,14} is formally possible but involves a more intricate constraint structure, including quaternion coordinates and additional first-class constraints. Work in this direction is in progress.
	
	Third, real two-dimensional materials with non-Coulombic screened potentials lie outside the exact closure class. An important question is whether a perturbative or variational extension of the amplitude-closure framework can describe such systems systematically, providing approximate but controlled corrections to the ideal Coulombic benchmark.
	
	Whether similar constructions can be extended to non-integrable systems remains an open and challenging question. For generic non-integrable potentials, the amplitude-closure condition will not hold exactly, and the relation between classical phase-space geometry and discrete spectral structure remains to be explored.

	%----------------------------------section 7--------------------------------
	\section{Conclusion}\label{sec:conclusion}
	
	This work provides an affirmative answer to the long-standing question of whether the full discrete bound-state spectrum of a non-trivial system can be derived exactly from classical mechanics without independent quantum axioms, at least for the integrable two-dimensional Coulomb problem. We have shown that the full bound-state spectral structure---energy levels, degeneracies, and the form of the linear evolution equation---can be derived exactly from classical Hamiltonian mechanics augmented by a single phase-scale parameter $\alpha$. No semiclassical approximation, no short-wavelength limit, and no independent quantum postulates are required.
	
	Our principal results are threefold. First, we have established the amplitude-closure criterion: a classical propagator kernel built from the two-point action and the Van Vleck amplitude satisfies a linear Schr\"odinger-type evolution equation if and only if the spatial Laplacian of the amplitude vanishes identically. Quadratic Hamiltonians form an exact closure class for which classical propagation and linear wave evolution are mathematically identical.
	
	Second, we have demonstrated that the Levi-Civita conformal regularization maps every negative-energy shell of the two-dimensional Coulomb problem onto an isotropic harmonic oscillator in image space, and that this mapping is structurally unique: the Coulomb potential is the only central potential that can be shell-wise mapped into a quadratic Hamiltonian via holomorphic conformal regularization.
	
	Third, we have derived the complete discrete bound-state spectrum, including the characteristic odd-integer Rydberg series, the $1/(2n-1)^2$ energy ratios, and the $N$-fold degeneracy structure, directly from the topology of the LC double cover and the integrability of the classical oscillator. The construction further determines the physical-space evolution operator uniquely, and resolves the self-adjoint domain ambiguity at the origin from purely geometric considerations.
	
	These results establish that for this integrable system, the discrete spectral structure is a direct consequence of classical phase-space geometry, with the linear wave equation serving as its natural linear representation. The framework provides a pure classical-geometry benchmark for interpreting exciton spectra in two-dimensional materials, and invites further investigation into the classical geometric origins of quantum spectral structure in integrable systems.

	% ==================== 致谢 ====================
	\textit{Acknowledgments}---We acknowledge the use of the Kimi large language model for formula derivation assistance and the DouBao large language model for English language polishing during the preparation of this manuscript.
	
	% ==================== 利益冲突声明 ====================
	The authors declare no competing financial interest.
	
	% ==================== 数据可用性声明 ====================
	\textit{Data availability}---There are no publicly available research data or software supporting this manuscript. Requests for further information or data should be sent to the authors.
	
	% ==================== 参考文献 ====================
	\bibliography{references}
	
\end{document}